\documentclass[conference]{IEEEtran}
\IEEEoverridecommandlockouts
\usepackage[utf8]{inputenc}
\usepackage{pgfplots}
\DeclareUnicodeCharacter{2212}{−}
\usepgfplotslibrary{groupplots,dateplot}
\usetikzlibrary{patterns,shapes.arrows}
\pgfplotsset{compat=newest}
\usepackage{standalone}
\usepackage{cite}
\usepackage{amsmath,amssymb,amsfonts}
\usepackage{algorithmic}
\usepackage{graphicx}
\usepackage{textcomp}
\usepackage{xcolor}
\usepackage{subcaption}
\usepackage{placeins}
\usepackage{xcolor}

\usepackage{hyperref}
\usepackage{booktabs}
\usepackage{siunitx}
\usepackage{float}
\newcommand{\ra}[1]{\renewcommand{\arraystretch}{#1}}

\newcommand{\update}[1]{{#1}}

\usepackage{xspace}
\newcommand{\zero}[0]{\gls{z3ro}\xspace}
\newcommand{\zerop}[0]{\zero precoder\xspace}
\newcommand{\mrt}[0]{\gls{mrt}\xspace}
\newcommand{\mrtp}[0]{\mrt precoder\xspace}
\newcommand{\pa}[0]{\gls{pa}\xspace}
\newcommand{\pas}[0]{\glspl{pa}\xspace}

\usepackage[acronym,shortcuts]{glossaries}
\makeglossaries
\newacronym[plural=BSs,firstplural=base stations (BSs)]{bs}{BS}{base station}
\newacronym[plural=PAs,firstplural=power amplifiers (PAs)]{pa}{PA}{power amplifier}
\newacronym{sdr}{SDR}{signal-to-distortion ratio}
\newacronym{snr}{SNR}{signal-to-noise ratio}
\newacronym{sndr}{SNDR}{signal-to-noise-and-distortion ratio}
\newacronym{snidr}{SNIDR}{signal-to-noise-and-interference-and-distortion ratio}
\newacronym{los}{LOS}{line-of-sight}
\newacronym{z3ro}{Z3RO}{zero third-order distortion}
\newacronym{mimo}{MIMO}{multiple input multiple output}
\newacronym{mrt}{MRT}{maximum ratio transmission}
\newacronym{dpd}{DPD}{digital pre-distortion}

\newacronym{ula}{ULA}{uniform linear array}

\newacronym{psd}{PSD}{power spectral density}
\newacronym{aclr}{ACLR}{adjacent channel leakage ratio}
\newacronym{cpu}{CPU}{central processing unit}
\newacronym{iid}{i.i.d.}{independently and identically distributed}
\newacronym{ue}{UE}{user equipment}
\newacronym{nlos}{NLoS}{non-line-of-sight}
\newacronym{hpbm}{HPBM}{half power beam width}
\newacronym{rts}{RTS}{ray tracing simulator}
\newacronym{ag}{AG}{array gain}
\newacronym{arp}{ARP}{Antenna Reference Point}
\newacronym{ecdf}{eCDF}{empirical cumulative distribution function}
\newacronym{evm}{EVM}{Error Vector Magnitude}
\newacronym{fft}{FFT}{fast Fourier transform}
\newacronym{ib}{IB}{in-band}
\newacronym{im}{IM}{intermodulation}
\newacronym{mf}{MF}{matched filtering}
\newacronym{mr}{MR}{maximum ratio}
\newacronym{ofdm}{OFDM}{orthogonal frequency division multiplexing}
\newacronym{oob}{OOB}{out-of-band}
\newacronym{papr}{PAPR}{peak-to-average power ratio}
\newacronym{rx}{RX}{Receiver}
\newacronym{trp}{TRP}{Transmission Reception Point}
\newacronym{tx}{TX}{transmitter}
\newacronym{zf}{ZF}{zero forcing}
\newacronym{dl}{DL}{downlink}
\newacronym{rzf}{RZF}{regularized zero forcing}
\newacronym{slc}{SLC}{spatial leakage suppression}
\newacronym{kpi}{KPI}{key performance indicator}
\newacronym{awgn}{AWGN}{additive white Gaussian noise}
\newacronym{qam}{QAM}{quadrature amplitude modulation}
\newacronym{amam}{AM/AM}{amplitude modulation to amplitude modulation}
\newacronym{ampm}{AM/PM}{amplitude modulation to phase modulation}

\newacronym{zmcscg}{ZMCSCG}{zero mean circularly symmetric complex Gaussian}

\def\BibTeX{{\rm B\kern-.05em{\sc i\kern-.025em b}\kern-.08em
    T\kern-.1667em\lower.7ex\hbox{E}\kern-.125emX}}
\begin{document}

\title{Measurement-Based Validation of Z3RO Precoder to Prevent Nonlinear Amplifier Distortion in Massive MIMO Systems}

\author{\IEEEauthorblockN{Thomas Feys, Gilles Callebaut, Liesbet Van der Perre, Fran\c{c}ois Rottenberg
	}
	\IEEEauthorblockA{KU Leuven, ESAT-WaveCore, Ghent Technology Campus, 9000 Ghent, Belgium
	}
}

\maketitle

\begin{abstract}
In \gls{mimo} systems, precoding allows the base station to spatially focus and multiplex signals towards each user. However, distortion introduced by \acrlong{pa} nonlinearities coherently combines in the same spatial directions when using a conventional precoder such as \gls{mrt}. This can strongly limit the user performance and moreover create unauthorized \gls{oob} emissions. In order to overcome this problem, the \gls{z3ro} precoder was recently introduced. This precoder constraints the third-order distortion at the user location to be zero. In this work, the performance of the \gls{z3ro} precoder is validated based on real-world channel measurement data. The results illustrate the reduction in distortion power at the UE locations: an average distortion reduction of \SI{6.03}{\decibel} in the worst-case single-user scenario and \SI{3.54}{\decibel} in the 2-user case at a back-off rate of \SI{-3}{\decibel}.
\end{abstract}


\section{Introduction}
\subsection{Problem Statement}
In current wireless cellular communication systems, the \gls{pa} is the main source of energy consumption~\cite{energy}. In order to achieve good communication performance, \glspl{pa} should be highly linear. However, there is a trade-off between power-efficiency and linearity of the PA~\cite{pa_for_wireless}. Consequently, when using a PA at an energy-efficient operating point, nonlinear distortion will arise. This nonlinear distortion is detrimental for the performance of the wireless system and can furthermore lead to unauthorized \gls{oob} emissions. This is especially the case in massive MIMO systems, where beamforming allows the base station to transmit the signal in a certain spatial direction. Recent studies have shown that the nonlinear distortion is not spatially spread out, but follows the dominant beamforming direction~\cite{distortion_beamformed, distortion_beamformed2}. This is witnessed in particular in situations with predominant transmission in one or a few directions (e.g. \gls{los} situations with only few users). This beamforming of the nonlinear distortion to the user location can strongly decrease the \gls{sdr} at the user location, which inherently limits the user performance. Additionally, this can introduce or strengthen unauthorized \gls{oob} emissions at both the user and unintended locations. 


\subsection{Conventional Solutions and Their Limitations}
One approach to avoid nonlinear distortion is to use the \gls{pa} with a certain back-off power, i.e., reduce the transmit power to set the operating point of the \gls{pa} sufficiently far away from the saturation point. This method avoids that the \gls{pa} is operating in the nonlinear region, hence avoiding nonlinear distortion. However, increasing the back-off, reduces the energy-efficiency of the \gls{pa}. Another drawback is the fact that reducing the transmit power can lead to poor \gls{snr} when the user experiences a high path loss.   

An alternative solution to the problem is to use \gls{dpd} techniques~\cite{pa_for_wireless}. These techniques pre-distort the unamplified signal to compensate for the nonlinear behavior of the \gls{pa}. This pre-distortion results in the amplified signal being a linear amplification of the original signal. Unfortunately, \gls{dpd} techniques introduce an additional complexity burden, which is especially present in massive MIMO systems where the \gls{dpd} has to be used on a per-antenna basis. Another limitation of \gls{dpd} is the fact that it can only account for weak nonlinear effects. Even a perfect \gls{dpd} can only linearize the \gls{pa} function up to the saturation point of the \gls{pa}. This is especially problematic for high \gls{papr} signals such as \gls{ofdm} signals~\cite{ofdm_papr}, which are typically used in current 5G massive MIMO systems. 

\update{More recent works have looked at how to reduce this nonlinear distortion by taking it into account in the precoding design~\cite{distortion-aware, distortion-cancellation, efficient_precoding_undr_pa_nonlins}. These works rely on iterative procedures to allow the \gls{pa} to work closer to saturation and/or to alleviate the need for \gls{dpd}. In~\cite{z3ro}, the Z3RO precoder was derived which circumvents the need for an iterative approach as a closed-form solution has been derived which maximizes the~\gls{snr} while nulling the third-order distortion term.}

\subsection{Contributions}
In~\cite{z3ro}, the authors proposed a novel precoder that allows the \gls{pa} to work close to saturation, while cancelling the third order nonlinear distortion at the user location. The \acrlong{z3ro} precoder is obtained by maximizing the \gls{snr}, while constraining the third-order distortion at the user location to zero. The precoder is obtained by saturating a subset of the antennas and inverting the sign of their phase shift. \update{In this study, the \zerop (originally designed for the single-user case) is expanded to the multi-user case. Additionally, given the novelty of the \zerop, measurement-based simulations are performed to further evaluate its performance.} The open-source channel measurements presented in~\cite{measurements} are used as the basis for a simulation-based assessment. Using these channel measurements, it is possible to calculate the distortion value at all measured locations. \update{Based on these simulations, the performance of the \zerop is assessed. A comparison against MRT is made, given that MRT is the optimal precoder for the single-user case when no distortion is present. Additionally, for the multi-user case only a small amount of users is considerd given that no interference-cancellation is present in the Z3RO and MRT precoders.}  

\update{\textbf{Notations}: Superscript $(.)^*$ stands for the conjugate operator. Subscripts $(.)_m$, $(.)_k$ and $(.)_l$ denote the antenna, user and location index respectively. The symbol $\mathbb{E}(.)$ denotes the expectation.}

\section{Transmission Model}
\subsection{Transmit Signal}
We consider a massive MIMO system where the \gls{bs} is equipped with $M$ antennas and $K$ single-antenna users are served using the same time and frequency resources. The complex transmit signal for user $k$, denoted as $s_k$, is assumed to be complex Gaussian distributed with variance $p_k$, and uncorrelated with the signal of other users. The signal is precoded at antenna $m$, using the coefficient $w_{m,k}$. The signal before the \pa is thus represented as 
$$x_{m} = \sum_{k=1}^K w_{m,k} s_k.$$ 
For the sake of clarity and without loss of generality, the \gls{pa} gain is set to one.
\subsection{Z3RO and MRT Precoders}
In \gls{mrt}, the complex precoding weights are matched to the channel coefficients in order to maximize the \gls{snr} at the intended users. The \zerop uses a number of saturated antennas $M_s$ to compensate for the distortion caused by all other antennas, improving the \gls{sdr} at the intended users.

The total power at the input of all the \pas is $p_T=\mathbb{E}\left(\sum_{m=0}^{M-1} |x_m|^2\right)=\sum_{k=0}^{K-1} p_k \sum_{m=0}^{M-1}|w_{m,k}|^2$. The average power at the input of each \pa is $p_{\mathrm{in}}=p_T/M$, with $p_{\mathrm{in}}=\sum_{k=0}^{K-1}p_k$. In order to respect this power budget, a normalization factor $\alpha_k$ ensures that $\sum_{m=0}^{M-1}|w_{m,k}|^2=M$. Hence, the back-off at the \pa becomes $p_{\mathrm{in}}/p_{\mathrm{sat}}$.

The classical \mrtp is defined as
\begin{align*}
    w_{m,k}^{\mathrm{MRT}}=\alpha_{k}^{\mathrm{MRT}} h_{m,k}^{*},
\end{align*}
where
\begin{align*}
\alpha_{k}^{\mathrm{MRT}}=\sqrt{\frac{M} {\sum_{m^{\prime}=0}^{M-1}\left|h_{m^{\prime},k}\right|^{2}}}.
\end{align*}

The \zerop~\cite{z3ro} is defined as
\begin{align*}
w_{m,k}^{\mathrm{Z} 3 \mathrm{RO}, M_{s}}=\alpha_k^{\mathrm{Z3RO}} h_{m,k}^{*}\left\{\begin{array}{ll}
-\gamma_k & \text { if } m=0, \ldots, M_{s}-1 \\
1 & \text { otherwise }
\end{array},\right.
\end{align*}
where $\gamma_k$ is the additional gain of the saturated antennas given by 
\begin{align*}
\gamma_k=\left(\frac{\sum_{m^{\prime}=M_{s}}^{M-1}\left|h_{m^{\prime},k}\right|^{4}}{\sum_{m^{\prime \prime}=0}^{M_{s}-1}\left|h_{m^{\prime \prime},k}\right|^{4}}\right)^{1 / 3}
\end{align*}

and $\alpha_k^{\mathrm{Z3RO}}$ is the power normalization constant defined as
\begin{align*}
\alpha_k^{\mathrm{Z3RO}}=\frac{\sqrt{M}}{\sqrt{\sum_{m^{\prime}=M_{s}}^{M-1}\left|h_{m^{\prime},k}\right|^{2}+\gamma^{2} \sum_{m=0}^{M_{s}-1}\left|h_{m,k}\right|^{2}}} .
\end{align*}
\update{The Z3RO precoder has a similar complexity as MRT with the main difference being the additional computation of $\gamma_k$. }

\subsection{PA Model}
In this study, the Rapp \pa model~\cite{rapp} is used. This model has a zero \gls{ampm} characteristic and an \gls{amam} characteristic which is modelled by
$$
y_{m}=\frac{x_{m}}{\left(1+\left|\frac{x_{m}}{\sqrt{p_{\mathrm{sat}}}}\right|^{2 S}\right)^{\frac{1}{2 S}}},
$$
where the amplified signal at antenna $m$ is denoted as $y_m$ and $S$ is the parameter which controls the smoothness of the transition from the linear region to the saturated region. $S=2$ is chosen for the assessment performed in this study. 

\subsection{Channel Measurements}
\begin{figure*}
    \centering
        \begin{subfigure}{0.90\columnwidth}
        \centering
            \includegraphics[width=0.8\columnwidth]{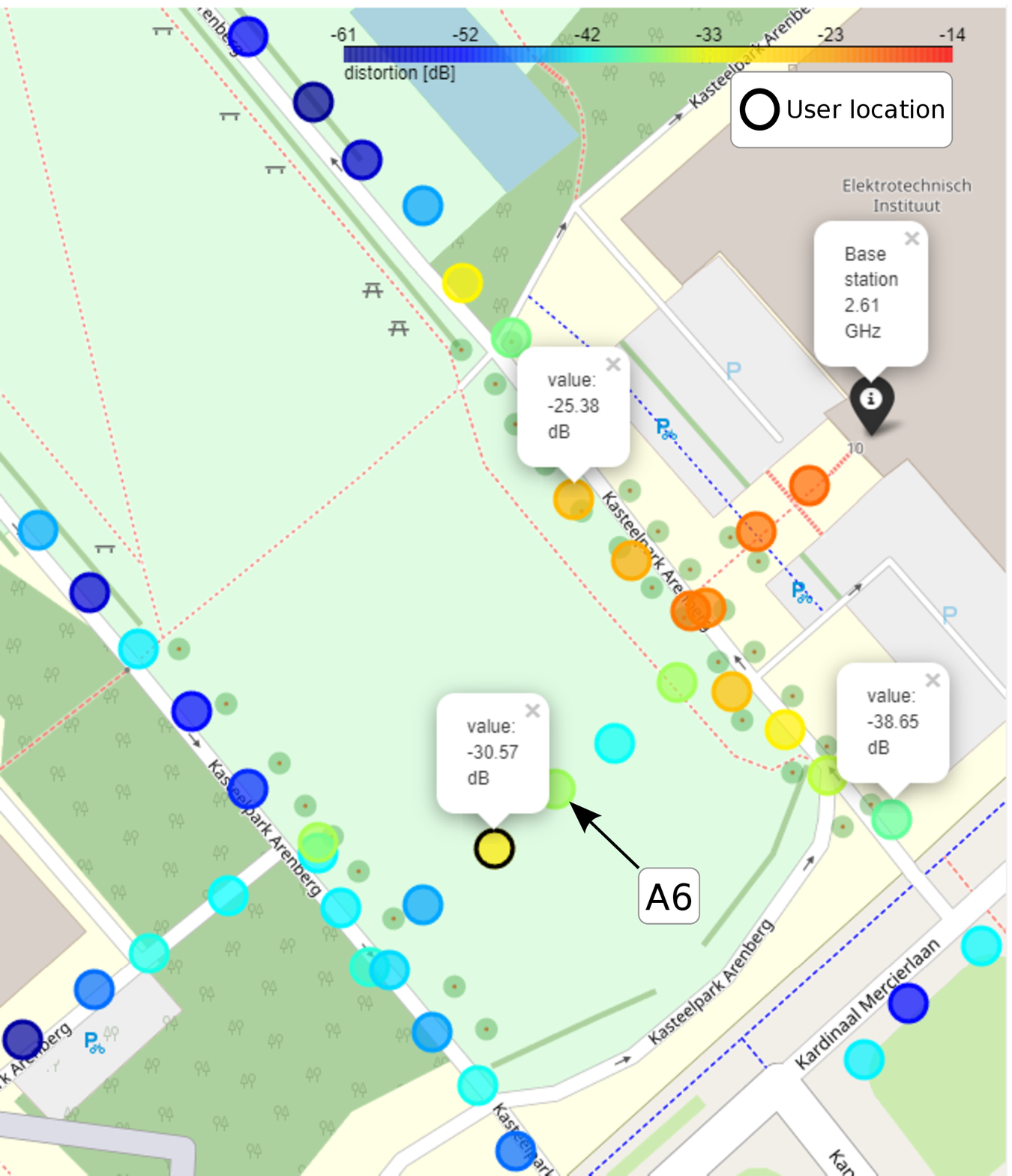}%
            \caption{Z3RO}%
            \label{subfiga}%
        \end{subfigure}\hspace*{35pt}%
        \begin{subfigure}{0.90\columnwidth}
        \centering
            \includegraphics[width=0.8\columnwidth]{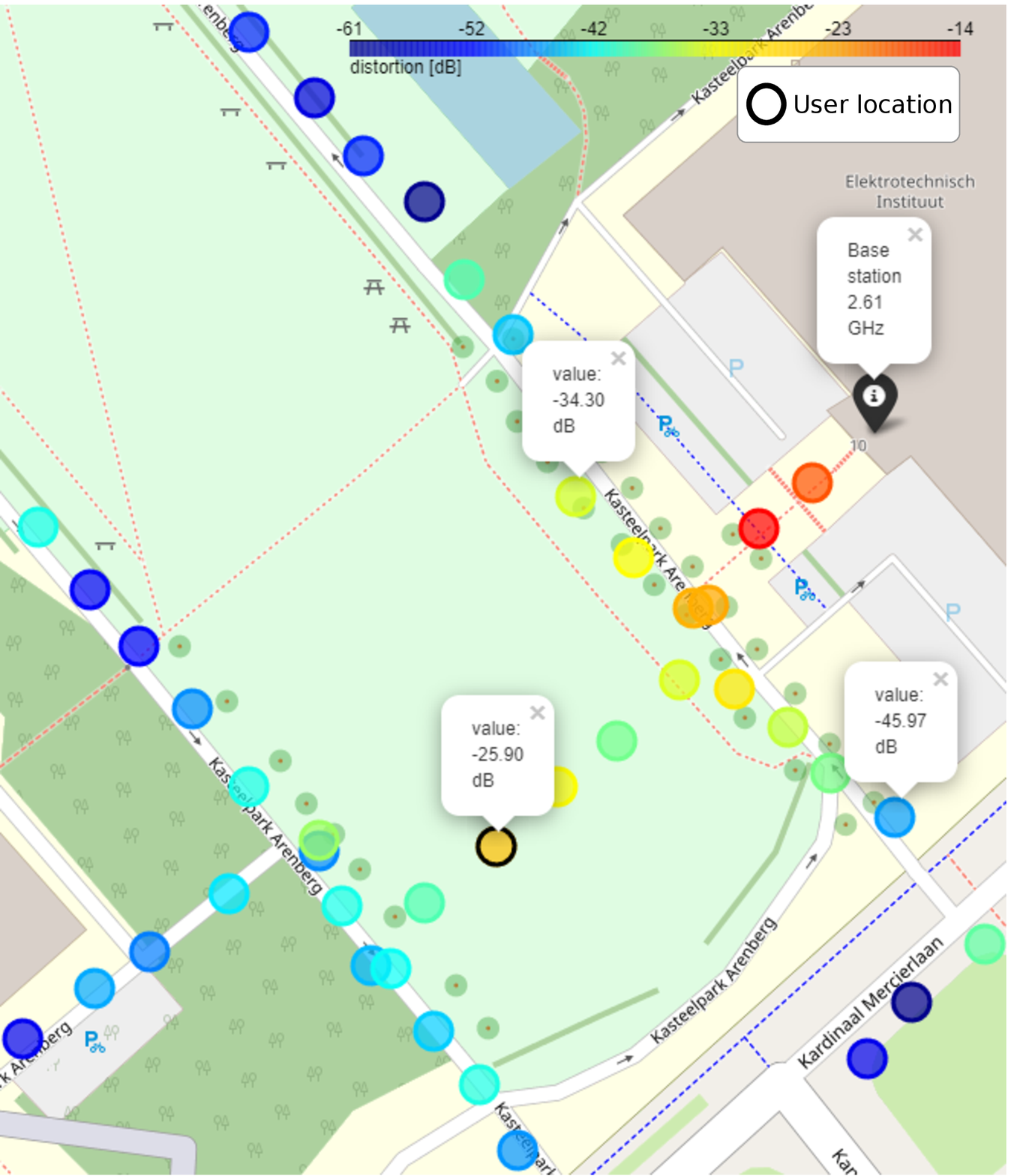}%
            \caption{MRT}%
            \label{subfigb}%
        \end{subfigure}\hfill%
    \caption{\small Heatmap of the distortion at all observer locations, for $p_{\mathrm{in}}/p_{\mathrm{sat}}$ = \SI{-3.1}{\decibel}. The distortion level at the user location is reduced for the \zerop, while it is slightly higher at unintended locations. \update{The circles denote the measurement locations, and the black icon the base station. The position highlighted with an arrow is the position used in Fig.~\ref{fig:r}. (approximate area of \SI{25000}{\metre\squared})}}
    \label{fig:single_user_heatmap}
\end{figure*}

In~\cite{measurements}, a measurement campaign was performed to capture real-world channel measurements. In this study, these open-source channel measurements\footnote{The open-source measurements are available here: \url{https://dramco.be/massive-mimo/measurement-selector/\#Sub-GHz}} are used to evaluate the \zerop. The details of the measurements are summarized in Table~\ref{tab:measurements}. These channel measurements are available for 42~observer~locations as illustrated in Fig.~\ref{subfiga}. The set of locations is defined as $\mathcal{L}$. The measured complex channel gain from antenna $m$ to an observer at location $l \in \mathcal{L}$ is denoted as $\tilde{h}_{m,l}$, while a complex channel gain from antenna $m$ to user $k$ is denoted as $h_{m,k}$. The observer location, i.e., unintended user location, is used to evaluate the distortion experienced at that specific position. This allows the study of the distortion level at both the user location and all unintended (observer) locations. 



\begin{table}[]
    \centering
    \ra{1.3}
    \caption{Overview of the open-source measurements.}
    \label{tab:measurements}
    \begin{tabular}{@{}lll@{}}
    \toprule
         Parameter & Symbol & Value  \\
    \midrule
         Carrier frequency & $f_c$ & 2.61 GHz\\
         Number of base station antennas & M & 32\\
         Base station Array configuration & & ULA\\
         Type of BS antenna & & Patch\\
         Type of UE antenna & & Dipole\\
         BS height & & 7 m\\
         UE height & & 1.5 m\\
         Number of measurement positions & & 42\\
    \bottomrule
    \end{tabular}
\end{table}

\subsection{Received Signal}
The received signal at observer location $l\in \mathcal{L}$ is given by
\begin{align*}
r_l = \sum_{m=0}^{M-1} \tilde{h}_{m,l} \cdot y_m + v_l ,
\end{align*}
where $v_l$ is zero mean complex Gaussian noise with variance $\sigma_v^2$ and $y_m$ is the amplified transmit signal, precoded to target one or several users.

\section{Performance Assessment}

\subsection{Performance Evaluation Metrics}\label{sec:kpi}
In order to objectively compare \zero precoding against \mrt precoding, the absolute distortion level (expressed in \si{\decibel}), is evaluated. Given that the absolute distortion level is correlated with the \gls{oob} radiation, it is a measure of the potential \gls{oob} leakage, and thus the \gls{aclr}. Consequently, studying the absolute distortion level allows us to get a better view on the \gls{oob} distortion level as well as the \gls{ib} distortion level (which impacts the \gls{sdr}).
The absolute distortion level at all observer locations can be evaluated using the following methodology. First, the received signal is decomposed in two parts: a desired part and an uncorrelated additive ``noise'' term following the Bussgang theorem~\cite{bussgang1952crosscorrelation,demir2020bussgang}. Thereafter, the \gls{sndr} and \gls{snidr} at the user location(s) are determined for the single and multi-user case.

\subsection{Single-User Case}
 In this section, a scenario is considered where a single-user is served by the base station. Given that having more users spatially spreads out the nonlinear distortion, the single-user scenario represents the worst-case in terms of the coherent combining of the nonlinear distortion at the user location. We define the user location index as $\tilde{l} \in \mathcal{L}$.
 
 Following the Bussgang \textit{decomposition}~\cite{demir2020bussgang}, the received signal for a single-user case can be decomposed as
\begin{align}\label{eq:busgang-single-user}
    r_l = \underbrace{G_l s}_\text{Desired signal} + \underbrace{d_l}_\text{Nonlinear distortion} + \underbrace{v_l}_{\text{Noise}},
\end{align}
where $l\in \mathcal{L}$, $d_l$ denotes the nonlinear distortion, that is uncorrelated with the transmit symbol $s$. $G_l$ is the linear gain that can be calculated as $G_l=\mathbb{E}\left(r_l s^{*}\right) / p$~\cite{demir2020bussgang}. The received signal variance is denoted as $|G_l|^2p$. The distortion variance is calculated using the knowledge that $s$, $v$ and $d_l$ are uncorrelated. As such, this leads to the following expression,
\begin{align}\label{eq:distortion-single-user}
\mathbb{E}\left(|d_{l} |^{2}\right)=\mathbb{E}\left(|r_{l} |^{2}\right)-|G_{l} |^{2} p-\sigma_{v}^{2}.
\end{align}
Using this expression it is possible to calculate the \gls{sndr} at the user location for which index $l=\tilde{l}$,
\begin{align*}
\mathrm{SNDR} = \frac{|G_{\tilde{l}}|^{2} p}{\mathbb{E}\left(|d_{\tilde{l}}|^{2}\right)+\sigma_{v_{\tilde{l}}}^{2}}.
\end{align*}

Next to the distortion level (\ref{eq:distortion-single-user}), the symbol rate at user k, expressed in bits/symbol is also considered. Assuming the worst-case of having the distortion Gaussian distributed, a lower bound on the achievable rate can be obtained as follows 
\begin{align}\label{eq:R-single-user}
R = \log_2(1+\mathrm{SNDR}).
\end{align}

\begin{figure}[H]
    \centering
    \includegraphics[clip, trim=2.5cm 2cm 2.5cm 3cm, scale=0.6]{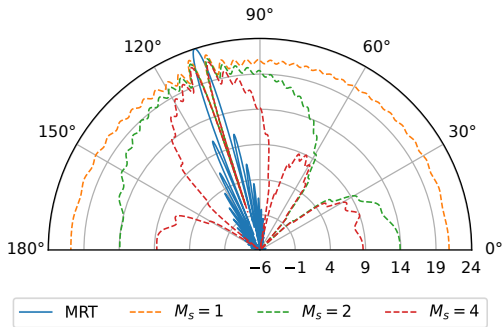}%
    \caption{\small Third-order distortion pattern of \mrt vs. \zero [\si{\decibel}], for $M=32$ (adopted from~\cite{z3ro}).}%
    \label{fig:radiation}
\end{figure}

\begin{figure}[!htb]
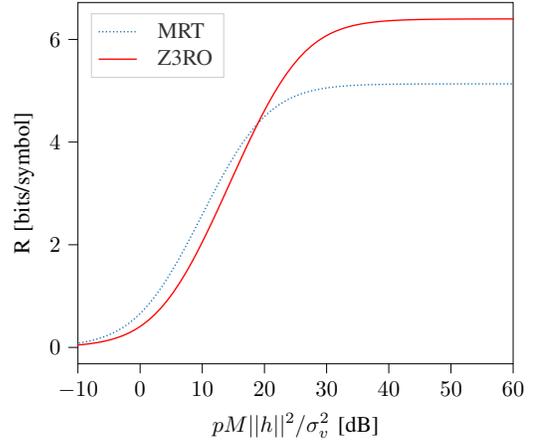
\centering
  \includestandalone[width=0.8\linewidth]{fig/location_A6}
  \caption{\small For a system that is noise limited, \mrt has a higher symbol rate. For a distortion limited system, \zero has a higher symbol rate. The considered user position is highlighted in Fig.~\ref{fig:single_user_heatmap}. ($p_{\mathrm{in}}/p_{\mathrm{sat}}=\SI{-3.1}{\decibel}$).}
  \label{fig:r}
\end{figure}

\begin{figure*}
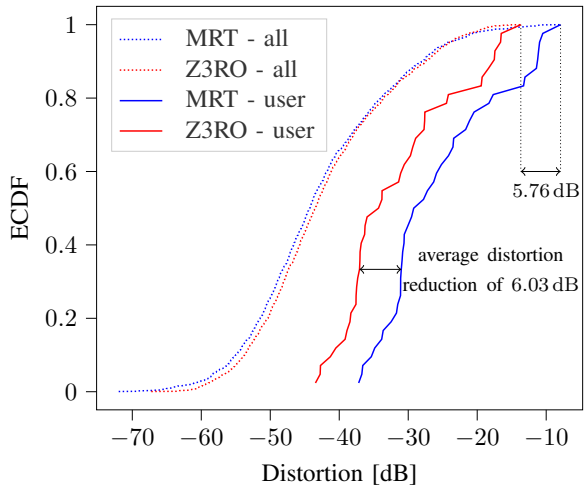
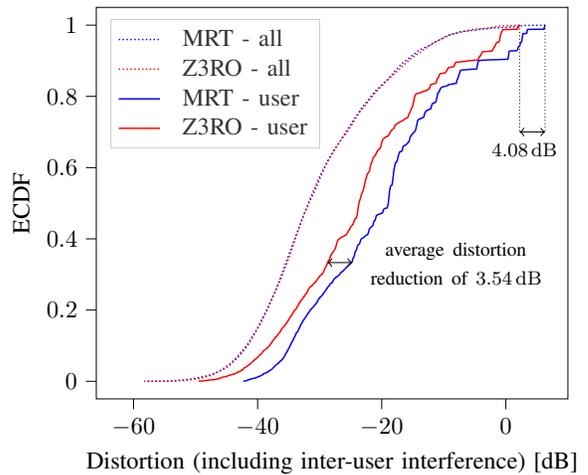

     \centering
     \begin{subfigure}[b]{0.48\linewidth}
         \centering
         \includestandalone[width=0.9\linewidth]{fig/single_user_cdf}
  \caption{\footnotesize Single-user case. ($M_s=2$)}
  \label{fig:single_user_cdf}
     \end{subfigure}
     \hfill
     \begin{subfigure}[b]{0.48\textwidth}
         \centering
         \includestandalone[width=0.9\linewidth]{fig/cdf_2user_correct}
  \caption{\footnotesize Two-user case. ($M_s=4$)}
  \label{fig:2_user_cdf}
     \end{subfigure}
        \caption{\small \Gls{ecdf} of the distortion level for the single and multi-user case. The  distortion level at all possible user locations (solid) and at all user and observer locations (dashed) for \mrt and \zero precoding. In both scenarios, the \zerop reduces both the distortion at the user location (solid) and the maximum experienced distortion at the observer locations (dashed). ($p_{\mathrm{in}}/p_{\mathrm{sat}}$ = \SI{-3.1}{\decibel})}
        \label{fig:distortion_cdf}
\end{figure*}

 Utilizing this methodology, the absolute distortion level, found in (\ref{eq:distortion-single-user}), is calculated at all observer locations. This is done at an input back-off of \SI{-3.1}{\decibel} and the number of saturated antennas for the \zerop is set to two ($M_s = 2$). As seen in Fig.~\ref{fig:single_user_heatmap}, the distortion at the user location is reduced for the \zerop, which is to be expected given that the \zerop nulls the third-order distortion term at the user location. Additionally, the distortion is spread out spatially, which results in higher distortion values for the \zerop at the observer locations. This is consistent with the theoretical results obtained in~\cite{z3ro}, as can be seen in Fig.~\ref{fig:radiation}. \update{The results illustrated in Fig.~\ref{fig:single_user_heatmap} solely depict one possible user location. However, given the 42 measurement locations, it is possible to place the user once at all of these 42 locations. For a thorough comparison of both precoders, the \gls{ecdf} when placing the user at every possible location is shown in Fig.~\ref{fig:single_user_cdf}.} This shows that the distortion level at the user location for the \zerop is significantly lower than for the \mrtp. On average, the \zerop reduces the distortion at the user location by \SI{6.03}{\decibel}. We can also observe that the overall distortion power for the \zerop is slightly higher compared to the \mrtp at the unintended locations, i.e., the observer positions, confirming the theoretical behaviour as derived in~\cite{z3ro}. Furthermore, it also demonstrates that the worst-case scenario is avoided, i.e., for high distortion levels the \zero precoding results in \SI{5.76}{\decibel} less distortion than \gls{mrt}. 
 Besides the absolute distortion level, the symbol rate ($R$) is also considered. Fig.~\ref{fig:r} shows the symbol rate when the noise variance ($\sigma_v^2$) is varied, given a constant $p_{\mathrm{in}}/p_{\mathrm{sat}}=\SI{-3.1}{\decibel}$. As can be seen in Fig.~\ref{fig:r}, if the system is noise limited (i.e., $pM||h||^2/\sigma_v^2 < 20 dB$)\footnote{Note that $||h||^2$ represents the squared L2-norm of the channel.}, \mrt shows a slightly larger symbol rate. However, when the system is distortion limited (i.e.,  $pM||h||^2/\sigma_v^2 > 20 dB$), the \zerop has a higher symbol rate. This confirms and validates the intended behaviour of the \zerop , as it improves performance when the system is distortion limited. 

\subsection{Two-User Case}
For the two-user case the number of saturated antennas is now equal to four ($M_s=4$). We define the user location index of user $k$ as $\tilde{l}_k \in \mathcal{L}$. Both users use the same time and frequency resources. The same reasoning used in the single-user case can be followed for the multi-user case. The signal, intended for user $k$, received at observer $l$ can be described as,
\begin{align}\label{eq:busgang-multi-user}
    r_{l,k} = \underbrace{G_{l,k}  s_k}_\text{Desired signal} + \underbrace{d_{l,k}}_{\substack{\text{Nonlinear distortion +} \\ \text{user interference}}} + \underbrace{v_{l,k}}_{\text{Noise}},
\end{align}
where $l\in \mathcal{L}$, $k=0,...,K-1$. Note that now $d_{l,k}$ also includes inter-user interference. The Bussgang gain is computed as  $G_{l,k}=\mathbb{E}\left(r_{l,k} s_k^{*}\right) / p_k$~\cite{demir2020bussgang}. The variance of the distortion and inter-user interference can be determined by,
\begin{align}\label{eq:distortion-multi-user}
\mathbb{E}\left(|d_{l,k} |^{2}\right)=\mathbb{E}\left(|r_{l,k} |^{2}\right)-|G_{l,k} |^{2} p_k-\sigma_{v_{l,k}}^{2}.
\end{align}

The \gls{snidr} at user $k$ is then given by 

\begin{align}\label{eq:snidr_k}
\mathrm{SNIDR}_k = \frac{|G_{\tilde{l}_k,k} |^{2} p_k}{\mathbb{E}\left(|d_{\tilde{l}_k,k}|^{2}\right)+\sigma_{v}^{2}}.
\end{align}


Equivalently to the single-user case, a lower bound on the achievable rate for the multi-user case can be obtained as follows, 
\begin{align}\label{eq:R-multi-user}
R_k = \log_2(1+\mathrm{SNIDR}_k).
\end{align}
The only difference in the multi-user case is that the achievable rate is also impacted by the inter-user interference.

\update{Utilizing equation~(\ref{eq:distortion-multi-user}), Fig.~\ref{fig:2_user_cdf} depicts the \gls{ecdf} of the distortion level (now including the inter-user interference) when considering every possible two-user combination given the 42 measurement locations. This figure shows an average reduction of the distortion power at the user locations of \SI{3.54}{\decibel}. The worst-case scenario, in terms of absolute distortion power, is also here reduced by \SI{4.08}{\decibel} for \zero precoding. This demonstrates the ability of the \zerop to reduce the distortion level at the user location, even in the two-user case. We note that the distortion reduction is smaller than in the single-user case. However, this is to be expected given that the two-user case is less stringent as the distortion is more spatially spread out when more users are present.}





\section{Conclusion}
In this study, the performance of the \zerop was assessed based on real-world channel measurements. The results demonstrate an average distortion reduction at the user location of \SI{6.03}{\decibel} in the most stringent single-user case and \SI{3.54}{\decibel} in the two-user case. Additionally, we showed that the lower-bound symbol rate for the \zerop is better as compared to \mrt, when the system is distortion limited. Finally, we confirmed that the \zerop introduces slightly more distortion at the unintended locations. However, we see that the worst-case scenario is avoided, as for high distortion levels, the \zero precoded signals experience \SI{5.76}{\decibel} less distortion than \gls{mrt}, in the single-user case and \SI{4.08}{\decibel} in the two-user case. This assessment validates the capability of the Z3RO precoder to operate massive MIMO systems closer to the saturation of the \glspl{pa} and hence increase the energy efficiency considerably. \update{In this study, promising results with respect to distortion mitigation were found for the single and two-user cases. However, currently no inter-user interference cancelation is present in the Z3RO precoder, making it challenging to deploy the precoder to scenarios with more users. Future research should aim to address this limitation.}


\bibliographystyle{IEEEtran}
\bibliography{IEEEabrv,mybib}

\vspace{12pt}

\end{document}